\pdfoutput=1
\documentclass[fleqn,10pt]{wlscirep}
\usepackage[utf8]{inputenc}
\usepackage[T1]{fontenc}
\usepackage{lineno,hyperref}
\usepackage{amsmath,graphicx,makecell}
\usepackage{enumitem}
\graphicspath{{./fig/}}
\usepackage{threeparttable}
\usepackage{booktabs}
\usepackage{algorithmic}
\usepackage[linesnumbered, ruled, vlined]{algorithm2e}
\usepackage{adjustbox}
\usepackage{booktabs}
\usepackage{amssymb}
\usepackage{wrapfig}
\usepackage{multicol}         
\usepackage{multirow}         
\usepackage{url}
\usepackage{tabularx}
\usepackage{xurl}

\title{How Much Demand Flexibility Could Have Spared Texas from the 2021 Outage?}

\author[1]{Dongqi Wu}
\author[1]{Xiangtian Zheng}
\author[1]{Ali Menati}
\author[2]{Lane Smith}
\author[3]{Bainan Xia}
\author[3]{Yixing Xu}
\author[1]{Chanan Singh}
\author[1,4,*]{Le Xie}

\affil[1]{Department of Electrical and Computer Engineering, Texas A\&M University, College Station, Texas, USA}
\affil[2]{Department of Electrical and Computer Engineering, University of Washington, Seattle, Washington, USA}
\affil[3]{Breakthrough Energy Sciences, Seattle, Washington, USA}
\affil[4]{Texas A\&M Energy Institute, College Station, Texas, USA}

\affil[*]{Corresponding author: le.xie@tamu.edu}

\begin{abstract}
The February 2021 Texas winter power outage has led to hundreds of deaths and billions of dollars in economic losses, largely due to the generation failure and record-breaking electric demand. In this paper, we study the scaling-up of demand flexibility as a means to avoid load shedding during such an extreme weather event. The three mechanisms considered are interruptible load, residential load rationing, and incentive-based demand response. By simulating on a synthetic but realistic large-scale Texas grid model along with demand flexibility modeling and electricity outage data, we identify portfolios of mixing mechanisms that exactly avoid outages, which a single mechanism may fail due to decaying marginal effects. We also reveal a complementary relationship between interruptible load and residential load rationing and find nonlinear impacts of incentive-based demand response on the efficacy of other mechanisms.

\end{abstract}

\begin{document}
\flushbottom
\maketitle
\thispagestyle{empty}

\section{Introduction}

Unprecedented winter storms that hit across Texas in February 2021 have caused hundreds of deaths and billions of dollars in economic loss due to large-scale generation outages and record-breaking electric demand. 
To understand the power outage event, plenty of studies in academic communities have provided insights from technical, social, and economic perspectives based on publicly available information.
Busby \textit{et al.}~\cite{busby2021cascading} provided a retrospective of root causes and summarized the financial and political impacts of the blackout. Clack \textit{et al.}~\cite{clack2021texas} reviewed renewable and natural gas resources in ERCOT and analyzed the financial repercussions due to the event.
King \textit{et al.}~\cite{king2021timeline} reviewed the timeline of the event and analyzed the factors that contributed to disruptions in the energy service.
Shaffer \textit{et al.}~\cite{shaffer2021demand} revealed the effects of increased susceptibility to extreme cold weather due to the electrification of space heating.
Zhang \textit{et al.}~\cite{zhang2022texas} conceptually classified the 2021 Texas power crisis as a new energy insufficiency-caused power crisis and revealed the underlying reasons for the power crisis based on quantitative simulations.
Wood \textit{et al.}~\cite{wood2021never} proposed a collection of policy update recommendations to strengthen the Texas power grid against future extreme weather events.
In November 2021, the Federal Energy Regulatory Commission (FERC), the North American Electric Reliability Corporation (NERC) and relevant regional entities jointly issued an official comprehensive report that reviewed the chronology of the blackout event and provided key recommendations on reliability and resilience enhancement.~\cite{ferc_report}

However, a challenging question remains unsolved for the broader energy research community: How can insights be provided to long-term electric power grid planning to improve system reliability against extreme weather conditions? Gruber~\textit{et al.}~\cite{gruber2021winterization} argued that large-scale winterization of power plants has economic efficiency with high risks considering the low frequency of extreme freezing events in Texas. Menati and Xie~\cite{menati2021}  analyzed the potential impacts of load rationing and proper sizing of energy storage for different scenarios to improve the reliability of the grid in extreme weather conditions. In our preliminary study, Wu \textit{et al.}~\cite{wu2021open} preliminarily quantified the performance of multiple technical solutions to mitigate power outages using the proposed open-source Texas grid model and dataset, including winterization of energy system facilities, interconnected high-voltage direct current (HVDC) lines, energy storage, and improved demand flexibility. Considering that demand flexibility has higher practicability in terms of the cost and development period, this paper is narrowed to evaluate the effects of demand flexibility, that is, the overall capability in electricity demand sectors of altering end-use consumption by contractual or voluntary mechanisms.



In this paper, we quantitatively assess the effects of multiple demand flexibility mechanisms on mitigating power outages. For demand flexibility modeling, we propose a general formulation of demand flexibility resources that can represent heterogeneous demand flexibility mechanisms of interest, including interruptible load, residential load rationing, and incentive-based demand response. To facilitate cross-disciplinary collaboration, we release a ready-to-use open-source package, including a calibrated Texas grid simulation model, a blackout-related dataset, and simulation code for what-if analysis of various demand response schemes. By simulating the synthetic model along with the blackout-related dataset, we find that, due to decaying marginal effects, single mechanisms can largely, but not completely, avoid power outages. For example, we find that by scaling interruptible load to 700\% or establishing a large-scale residential load rationing program that curtails residential demand by 27\%, 75\% of forced load shedding could have been avoided. Therefore, we identify portfolios of mixing mechanisms that completely avoid the power outage. Additionally, we reveal a complex interaction between multiple mechanisms, that is, interruptible load and residential load rationing always have a complementary relationship, while incentive-based demand response may have counterintuitive nonlinear impacts on the efficacy of the other mechanisms.
\section{Formulation of Demand Flexibility}

With more frequent extreme weather events, energy portfolio transition towards decarbonization, and deepening penetration of intermittent renewables, come unique challenges for the power grid supply-demand balancing and reliability management. Although a large body of research has focused on scheduling dispatchable generation resources,~\cite{kazarlis1996genetic,gaing2003particle,coelho2006combining} there is an increasing interest in exploiting demand-side flexibility to address energy intermittency and scarcity.~\cite{shariatzadeh2015demand,wang2015load,jackson2021building, osti_1785329}
To assess the effects of demand flexibility against the February 2021 Texas winter storm, we focus on three mechanisms that represent most of the demand flexibility in the power market. They are interruptible load, residential load rationing, and incentive-based demand response (see more details in Section~\ref{section:data,model,simulation}). As of today, these three mechanisms account for more than 94\% of the demand flexibility in the ERCOT market~\cite{drercot} (see Appendix A. 1).
Interruptible load refers to a type of contractual demand response program that can interrupt large commercial and industrial loads to an arbitrary extent within a short period upon request, such as existing demand-side reserves in ERCOT.
Researchers have investigated the modeling~\cite{AALAMI2010243,1546842}, management~\cite{aminifar2009unit, huang2004integrating,yang2015stochastic}, and market design~\cite{bhattacharya2003competitive} of interruptible load.
Residential load rationing refers to a type of contractual demand response program that can restrict residential electricity consumption during energy scarcity, which is potentially implemented by observing energy consumption at the appliance level recorded by smart sensors. Electricity rationing has been widely used as a last resort to keep critical demands online during emergency conditions where severe electricity supply deficiencies are inevitable.~\cite{ROCHASOUZA2007296, DENOOIJ2009342} Note that although both interruptible load and residential load rationing are firm demand response programs with no uncertainty, being fulfilled on different load sectors may cause distinct impacts.
Incentive-based demand response refers to a type of voluntary demand response programs that can exploit the elasticity of residential loads by impelling voluntary and stochastic load shifting in response to EnergyCoupon-like bonus incentives~\cite{zhong2012coupon,xia2017energycoupon,ming2020prediction} rather than price-based incentives. EnergyCoupon-like bonus incentives encourage load reduction or shifting by some lottery-like mechanism, while price-based incentives implement it by changing the real-time price in the period of electricity crisis. It is worth noting that a price-based incentive aimed at reducing or shifting consumer load is not effective during this type of power outage event because wholesale market nodal prices reach their maximum and individual consumer utility changes during an emergency.

Despite the heterogeneous forms, we describe demand flexibility from three technical aspects that depend on control functions, including response time, duration, and uncertainty.~\cite{akrami2019power} Consequently, we define several key attributes, including ramping rate $R_t$ at time $t$, reduced load power $P_t$ at time $t$, and actual activation duration $T$. We propose a general formulation of demand flexibility sources as follows,
\begin{subequations}\label{eq:DR_def}
\begin{align}
    &R_{\text{min}}\leq R_t \leq R_{\text{max}},\\
    &\frac{dP_t}{dt}=R_t,\\
    &0\leq P_t \leq P_{\text{max}},\\
    &T_{\text{min}}\leq T \leq T_{\text{max}},
\end{align}
\end{subequations}
where $P_{\text{max}}$ represents the maximum capacity provided by the demand flexibility source, $R_{\text{max}}$ represents the maximum response rate, the absolute value of $R_{\text{min}}$ represents the maximum restoration rate, $T_{\text{min}}$ and $T_{\text{max}}$ represent minimum and maximum activation duration, respectively. Specifically,  $P_{\text{max}}$ is constant for deterministic demand response sources, while it can be a random variable following some distribution $\mathop{\mathbb{P}}$ in stochastic cases. Please see more details of implementation in Methods.

\section{Data, Model and Simulation Setup}\label{section:data,model,simulation}

\subsection{Electricity Demand and Generation Outage Data}
We aggregate actual load, actual generation, seven-day-ahead load forecast, and seven-day-ahead solar generation forecast data from ERCOT,~\cite{ercot_gridinfo} collect generation unit outage data from ERCOT,~\cite{ercot_generator_outage2} and obtain installed generation capacity data from EIA.~\cite{eia_capacity} It is worth mentioning that only part of all generation outages are included in the collected generation units outage data because resource entities may not disclose such information and outages shorter than two hours may not be included. For detailed profiling of various load sectors, we obtain load profile data for all profile types and weather zones from ERCOT~\cite{ERCOT_load_profiling} and aggregate demographic data from the U.S. Department of Housing and Urban Development~\cite{hud_data} and the U.S. Census Bureau.~\cite{census_data} For information on existing demand response programs, we aggregate actual load reduction time series that were caused by demand response programs from ERCOT.~\cite{ERCOT_interruptible_load}
Besides, we define counterfactual load data as the seven-day-ahead load forecast, counterfactual solar generation capacity as the seven-day-ahead solar generation forecast, and load shedding as the gap between the actual and counterfactual load data.

Following data collection, we estimate \textit{counterfactual wind generation} as described in Xu \textit{et al.}~\cite{BTE1} based on weather data.~\cite{wind_weather_data} Then we calculate the total available generation capacity along the timeline based on installed generation capacity data, generation unit outage data, counterfactual solar generation capacity, and counterfactual wind generation capacity. To improve the fidelity of our simulation and model different demand response proposals in richer detail, we further decompose the ERCOT weather-zone level total load data into three sectors: residential, business, and other. We estimate a varying percentage share of all sectors in the total load of each weather zone at each hour in February 2021 (see more details in Appendix A. 1).

\subsection{Synthetic Texas Electricity Grid Calibration}
Here we briefly describe the model calibration of original synthetic 2,000-node Texas grid (see the topology of the synthetic grid in Appendix A. 3). More details can be found in our previous studies.~\cite{wu2021open,breakthrough_data} We rigorously calibrate the model in several aspects: system topology, transmission lines rating, generation units capacity, and geographical demand distribution. The topology of the synthetic Texas 2,000-bus grid was originally presented in Birchfield \textit{et al.}~\cite{birchfield2017grid} Transmission line capacity ratings and impedance have been updated to properly incorporate the 2021 generation mix and demand distribution. The generator capacities have been updated to incorporate the installed generation capacity data~\cite{eia_capacity} from 2016 to 2021. Weather zone loads have been updated to reflect current data.~\cite{ercot_gridinfo} Additionally, we decompose the loads into multiple sectors utilizing the load profiling data, which can be used for the allocation of load shedding.

\subsection{Simulation Designs of the 2021 Blackout Event}
The overall idea of event reproduction is to successively perform load shedding and restoration to make the DCOPF converge at every moment with the \textit{total available generation capacity} and \textit{counterfactual load} time series as initial input.
Based on the calibrated simulation model and associated blackout event data, we propose the procedure of load shedding and restoration in Algorithm~\ref{algo:ls}, where $P_{\text{l0}}^t$, $P_{\text{g}}^t$, and $P_{\text{r}}^t$ represent the counterfactual load vector, generation vector, and total reserve at time $t$, respectively, while ${flg}_{\text{cvg}}$ indicates the convergence of the DCOPF algorithm, and $P_{\text{r,min}}$ represents the minimal system reserve to avoid system-wide collapse, which is 2,300 MW according to public ERCOT documents.\cite{nodal_operation_guides} 

\begin{algorithm}[htpb!]
\caption{Simulation process of reproducing the 2021 Texas blackout event}
\label{algo:ls}
\begin{algorithmic}
\FOR{$t=1,\,...\,,\,T$}
\STATE \textbf{Input}  $ P_{\text{l0}}^t$, $P_{g}^t$
\STATE $ P_{\text{l}}^t \longleftarrow P_{\text{l0}}^t - P_{\text{ls}}^{t-1}$
\STATE $P_{\text{r}}^t \longleftarrow \sum_i (P_{\text{g},i}^t - P_{\text{l},i}^t) + P_{\text{ls}}^{t-1},\,$ $\forall i$
\STATE ${flg}_{\text{cvg}} \longleftarrow \text{DCOPF}(P_{\text{g}}^t,P_{\text{l}}^t)$
\IF{NOT ${flg}_{\text{cvg}}$ or $P_{r}^t < P_{\text{r,min}}$}
\WHILE{NOT ${flg}_{\text{cvg}}$ or $P_{r}^t < P_{\text{r,min}}$}
\STATE $P_{\text{ls}}^t \longleftarrow P_{\text{ls}}^t + \Delta P_{\text{ls}}$
\STATE $P_{\text{r}}^t \longleftarrow P_{\text{r}}^t + \Delta P_{\text{ls}}$
\STATE ${flg}_{\text{cvg}} \longleftarrow \text{DCOPF}(P_{\text{g}}^t,P_{\text{l}}^t)$
\ENDWHILE
\ELSIF{$P_{\text{r}}^t > P_{\text{r,min}}$ and $P_{\text{ls}}^{t} > 0$}
\WHILE{${flg}_{\text{cvg}} \,\&\, P_{\text{r}}^t > P_{r,\text{min}} \,\&\, P_{\text{ls}}^t > 0$}
\STATE $P_{\text{ls}}^t \longleftarrow P_{\text{ls}}^t - \Delta P_{\text{ls}}$
\STATE $P_{\text{r}}^t \longleftarrow  P_{\text{r}}^t - \Delta P_{\text{ls}}$
\STATE ${flg}_{\text{cvg}} \longleftarrow \text{DCOPF}(P_{\text{g}}^t,P_{\text{l}}^t,P_{\text{ls}}^t)$
\ENDWHILE
\ENDIF
\ENDFOR
\end{algorithmic}
\end{algorithm}

\subsection{Demand Response Program Implementation}\label{appendix_B}
To add demand flexibility into the iterative load shedding algorithm (Algorithm~\ref{algo:ls}), we implement demand flexibility under each mechanism following the proposed general formulation presented in Equation~\eqref{eq:DR_def} with the key parameters defined in Table~\ref{tab:DR_parameters}. 

\begin{table*}[h]
    \centering
    \caption{Parameters of heterogeneous demand response programs}
    \begin{tabular}{lccc} \toprule
         & \begin{tabular}[c]{@{}l@{}}Interruptible load\end{tabular} 
         & \begin{tabular}[c]{@{}l@{}}Residential load rationing\end{tabular} 
         & \begin{tabular}[c]{@{}l@{}}Incentive-based demand response\end{tabular}    \\ \midrule 
        $R_\text{min}$ & -100\% per hour  & -10\% per hour & -100\% per hour \\
        $R_\text{max}$  & 50\% per hour & 10\% per hour  & 100\% per hour \\
        $P_\text{max}$  & Scaled-up ERCOT data  & 50\% & Random sampling \\
        $T_\text{max}$  & $\infty$  &  $\infty$ & 1 hour \\
        \bottomrule
    \end{tabular}
    \label{tab:DR_parameters}
\end{table*}

\paragraph{\textbf{Interruptible load}} We obtain the real historical ERCOT load reduction record from a presentation report,~\cite{ercotlr} including load components of the Emergency Response Service (ERS) and Responsive Reserve Service (RRS), price response, and demand response programs of utility companies. The historical record of interruptible load commitment for each hour defines the maximum available reduction from this category, as we assume that during the blackout event grid operators at each level have already exhausted their resources. In the simulation workflow, interruptible load is activated when either of the following conditions is met: (i) the operation reserve (sum of available generation minus sum of load) of the entire grid is less than 3000 MW; or (ii) there is ongoing forced load shedding or load rationing that makes the online demand less than the counterfactual demand. Once activated, the real power demand of load buses across the system is iteratively reduced until the total reduction exceeds the maximum available interruptible load resources capacity or the system reserve is back up above 3000 MW. The total reduction is distributed among the load buses based on their share of commercial or industrial components as determined in the load profiling and population data. In hypothetical scenarios where we increase the amount of available interruptable load, the maximum capacity is scaled up accordingly. It is worth noting that interruptible loads are typically firm loads of constant power consumption; therefore, the scaling-up rate, instead of the coverage rate, is used to indicate the scale.

\paragraph{\textbf{Residential load rationing}} We propose residential load rationing as a substitute for rotating outages during similar extreme events when there is a large deficiency in the electric power supply. The idea is to allow every household to retain access to the minimum amount of capacity that is necessary for basic life support and safety, while not overloading the bulk power grid under emergency conditions. In our model, a maximum percentage of load rationing is declared prior to simulation as an input parameter. In each hourly snapshot of the simulation, when the system cannot supply the load level (even after available interruptible load has been dispatched) or there is ongoing forced load shedding, the load rationing process is initiated. The process of determining the level of load rationing is similar to iterative load shedding in Algorithm~\ref{algo:ls}, except that the level of commitment is measured by an active percentage between zero and the maximum allowed rationing percentage. The active percentage means the percentage of all residential components of loads that are dropped for rationing (that is, 15\% active rationing means that all residential load can only get 15\% of their counterfactual capacity). The amount of residential demand at each load bus is determined by the ERCOT load profile data and the population size in the geographic zone they represent. Depending on the level of supply deficiency and the pattern of network congestion, a reduction in the maximum allowed rationing percentage might not be necessary for all hours.

\paragraph{\textbf{Incentive-based demand response}} We built the incentive-based demand response model based on data from the real-world EnergyCoupon experiment.~\cite{zhong2012coupon,xia2017energycoupon,ming2020prediction} The EnergyCoupon project attempted to alter the customer consumption pattern during peak load hours by giving coupons that can be used to participate in lotteries to win gift cards. The data obtained in that experiment are used to model user behavior in power consumption reduction when incentives are given. Users in the treatment group were classified into two groups: inactive, whose electricity consumption was very minimally impacted by the incentive; and active, whose consumption reduction was much more significant compared to the inactive group. We model the available capacity for load reduction from incentive-based demand response programs based on the characterized behavior of both groups, as well as the ratio between active and inactive participants. We simulate the effect of an incentive-based demand response program in hypothetical scenarios in which a percentage of residential customers participate in such programs. During the simulation, participants receive incentive signals at hours when \textit{additional} demand reduction is needed. If the system is already feasible using other forms of reduction (including forced load shedding) that are inherited from the past hours, no incentive will be given. The amount of reduction for each load bus is sampled from a random distribution derived on the basis of the data collected in the EnergyCoupon experiment. Also, if the load is already under reduction caused by either load rationing or forced load shedding, the reduction from incentives is discounted because customers are less likely to curtail as much when they already lost a portion of their load capacity.

\paragraph{\textbf{Priority of Different Mechanisms}} In the case of demand response portfolios, we define the priority in the following order: incentive-based demand response, interruptible load, and residential load rationing. In reality, the availability of each type of demand response must be procured in advance, and their activation depends on different conditions on the real-time grid supply and demand balance status. For grid operators that manage the demand-side resources, when the supply deficiency becomes large enough to trigger demand response mechanisms, similar to the allocation of power generation, demand response resources should also be prioritized based on their relative cost, response speed, and reliability. For example, interruptible load resources are usually long-term commitment-based contracts that are settled long before an extreme weather event. Gaining load reduction from such mechanism is almost guaranteed and requires little additional investment. In contrast, incentive-based voluntary reduction can have much more uncertainty in terms of how much MW of demand is available for curtailment, and initiators of such a program need to pay potential participants on the spot as the incentive. Based on the characteristic of these demand response programs, it would be desirable for utilities to execute all interruptible load contracts before starting any additional incentive-based programs.
In our simulation, the priorities of demand response models are reflected in a similar principle: in case the total generation capacity fails to meet the total or regional demand in the system, the interruptible loads are activated first to bring down the demand. If the reduction from available interruptible load is not enough, load rationing is activated up to a level determined by the current commitment percentage and the ramping limit. If the gap is not yet filled, incentive-based program participants receive a signal to reduce. Any further curtailment required to obtain a feasible solution is considered forced load shedding. Note that the activated reduction percentage from interruptible load and load rationing is preserved across hours, while the incentive-based demand response only lasts one hour and is only activated if the other reduction inherited from the past hours is not enough.


\section{Insights into the 2021 Texas Power Outage}
\subsection{Representation of the Texas Electric Power Grid in the 2021 Winter Storm}

\begin{figure*}[h]
	\centering
	\includegraphics[width=0.8\textwidth]{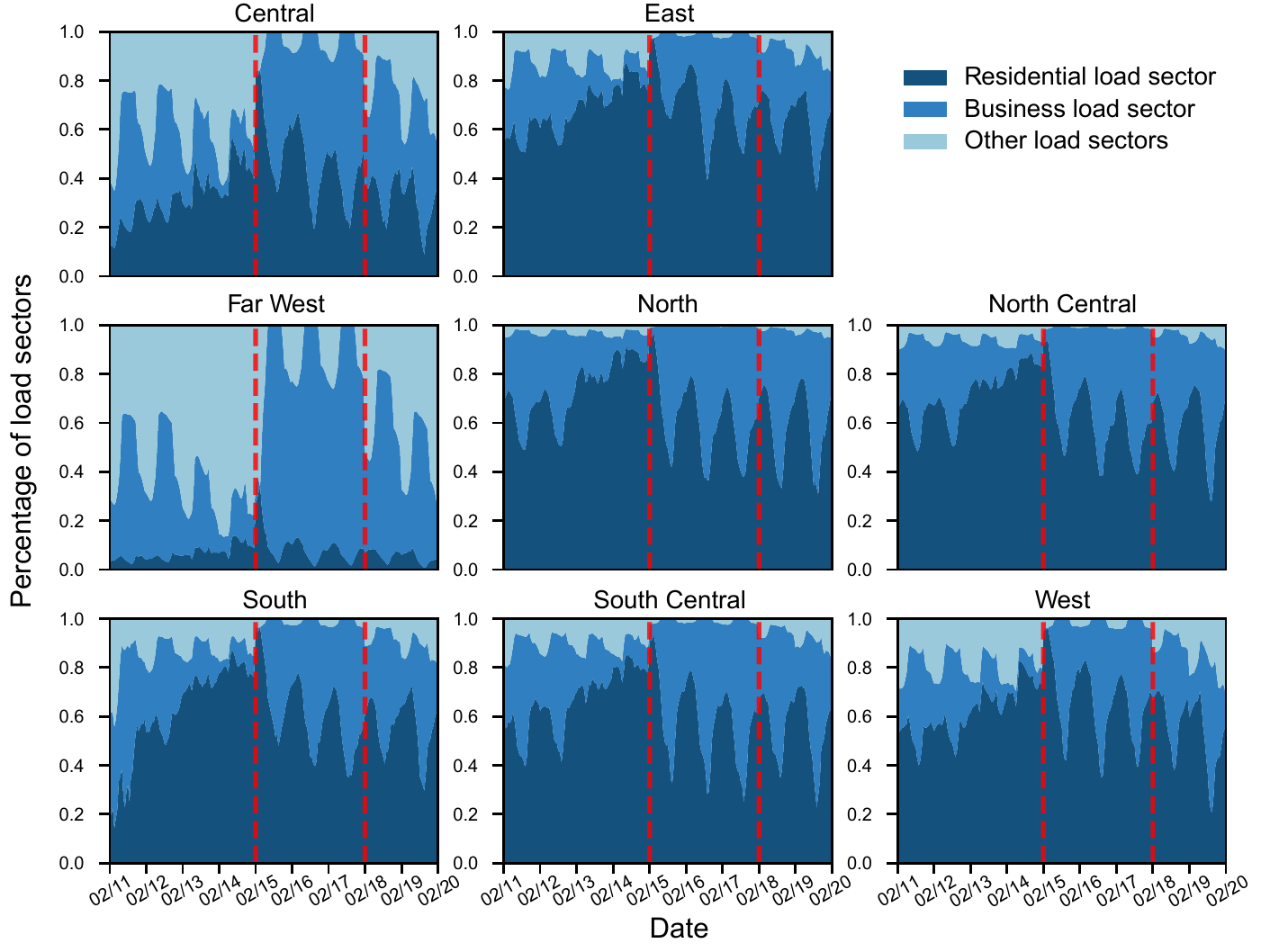}
	\caption{Visualization of load profiling over weather zones. The stacked plot presents the estimated load percentages of three load sectors in weather zones, which shows the significant rising proportion of residential load sectors with the winter storm incoming (starting from February 11). The dashed red vertical lines indicate the beginning and end of the blackout event.}
	\label{fig:sector_share}
\end{figure*}

We have developed a high-fidelity synthetic 2,000-node model of the Texas electric power grid that is tuned to represent its status around February 2021. This model was used for our previous study~\cite{wu2021open} on simulating the timeline of power outages from the Texas electricity crisis. In this study, we closely examine the effectiveness of various up-scaled demand flexibility mechanisms in mitigating the severity of the power outage by incorporating data and models that are more detailed. We improve the granularity of the electric demand model by decomposing load profile data according to residential, business, and other electric demand (see Methods); these designations are used for the modeling and implementation of different demand response programs.

Based on estimated load profiling data, we observed a significant increase in the proportion of residential load sectors with the incoming winter storm (Fig.~\ref{fig:sector_share}). Under normal conditions, residential load sectors should have only likely accounted for between 10\% and 50\% of the system electric demand. In particular, the distinct pattern in the Far West is mainly due to its sparse population. It is worth noting that load profiling is critical for demand flexibility assessment because different demand flexibility mechanisms may have distinct performance based on the real-time electricity demand of different load sectors under different scenarios.

\subsection{Reproduction of the 2021 Texas Power Outage}
We reproduce the 2021 Texas blackout event from February 15 to February 18 with limited demand flexibility by simulation based on the synthetic grid model, of which the main steps consist of forced load shedding, demand flexibility resource activation, and load restoration (see Methods). Here, we use Energy Not Served (ENS), a power system reliability index, as the metric to quantify the severity of the blackout event.~\cite{singh2018electric} ENS refers to cumulative forced load shedding and excludes the load reduction caused by demand flexibility.

\begin{figure}[h]
	\centering
	\includegraphics[width=0.5\columnwidth]{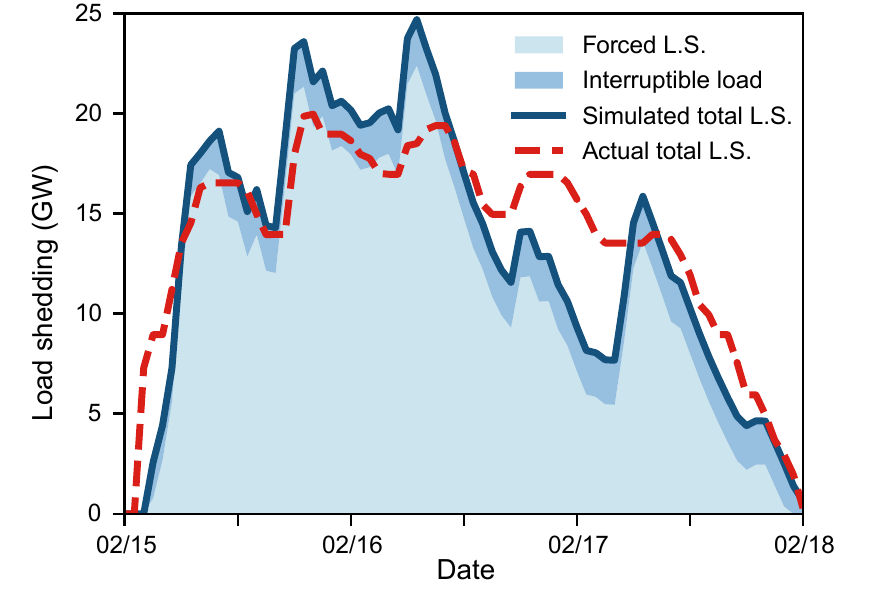}
	\caption{Simulated load shedding curve with anatomy of components in comparison with the actual total shedding. Here L.S. means load shedding. Simulated total load shedding equals to the sum of simulated interruptible load and forced load shedding.}
	\label{fig:baseline_loadshedding}
\end{figure}

We demonstrate the fidelity of the simulation model and the designed load shedding algorithm by the good match between the actual and simulated total load shedding (Fig.~\ref{fig:baseline_loadshedding}) that represent a total of 998.8 GWh and 955.5 GWh ENS (4.33\% difference), respectively. The correlation coefficient between the two time series is 0.88, which also indicates a good match between the simulated load-shedding process and the actual ERCOT record. The unavoidable mismatch is due to the combined effects of granularity difference between the synthetic model and real ERCOT grid, and exact system operation decisions under emergency conditions that are now disclosed to the general public.
The result also shows that the limited demand flexibility in the real-world grid has trivial effects on the mitigation of power outages under such extreme weather conditions, stressing the need to facilitate larger demand flexibility portfolios.

\section{Effects of Demand Flexibility against Extreme Weather}
\subsection{Effects of Demand Flexibility under a Single Mechanism}
We first evaluate the effects of demand flexibility under each individual demand response mechanism on varying scales, which are implemented using the general formulation (Equation~\eqref{eq:DR_def}). We determine the total activated size in real time of the demand response with the constraints of restoration and ramping rates before performing forced load shedding at each time step and then allocate the total activated size to the corresponding load sectors (see Section 3).

\begin{figure*}[h]
	\centering
	\includegraphics[width=1.0\textwidth]{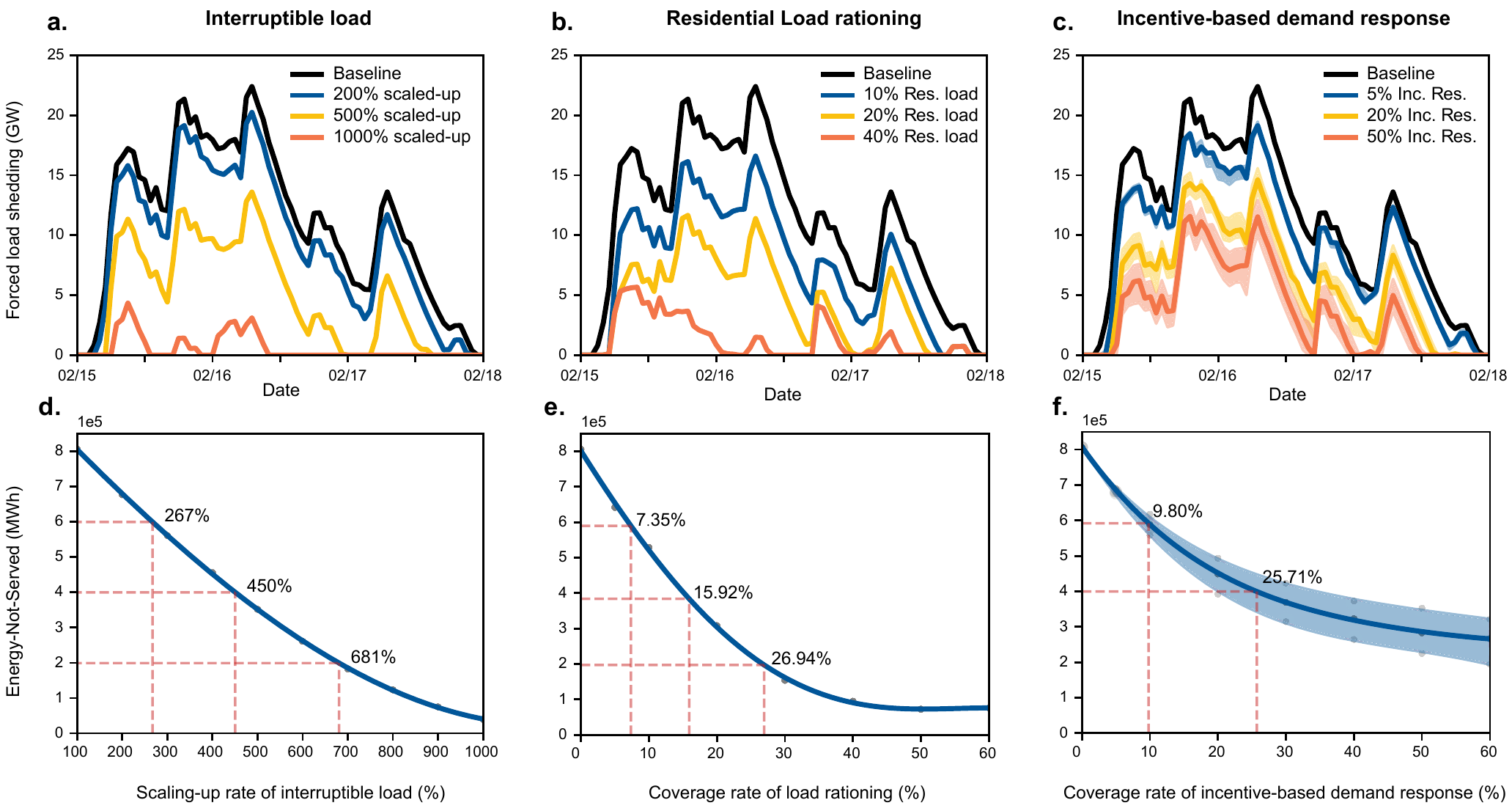}
	\caption{Forced load shedding curves considering the implementation of demand flexibility of various mechanisms and scales. Numbers in percentage in the figures mean the scale of demand response schemes such that 25\%, 50\%, and 75\% forced power outage can be avoided. a-c, The forced load shedding curves correspond to interruptible load, residential load rationing, and incentive-based demand response, respectively, where Res. load represents residential load and Inc. Res. represents incentivized residences. d-f, The curves of ENS (cumulative forced load shedding) are a function of the scale of the demand response program. The shadow black points represent actual simulation results, while the blue solid curves and bands correspond to regression results.}
	\label{fig:single_assess}
\end{figure*}

We quantify the effect of each individual mechanism by simulating demand response programs of each type and different level of scales (Fig.~\ref{fig:single_assess} a-c); the efficacy is assessed in terms of ENS (Fig.~\ref{fig:single_assess} d-f and Table \ref{tab:single}). It is worth noting that (i) we use the scaling-up rate instead of the coverage rate to represent the scale of interruptible loads that are typically firm loads with constant power consumption; and (ii) the shadow bands of incentive-based demand response in Figs.~\ref{fig:single_assess}-c and~\ref{fig:single_assess}-f are attributed to stochastic response in scenarios where customers may have more or less willingness than expected to coupon incentives. All single mechanisms exhibit decaying marginal effects along with scale, especially for residential load rationing and incentive-based demand response, stressing the need to balance system reliability and economics. This is because when the scale of a demand flexibility scheme is sufficiently large, an additional one-unit scale is only effective during a short period of peak forced load shedding, resulting in a lower per-unit contribution to the ENS reduction.
\begin{table}[h]
\centering

\begin{adjustbox}{width=\textwidth}
\begin{tabular}{llcccccc} \toprule
        \multirow{2}{*}{\begin{tabular}[c]{@{}l@{}}Interruptible load\end{tabular}} 
        & Size
         & Baseline & 200\% scaled-up & 400\% scaled-up & 600\% scaled-up & 800\% scaled-up & 1000\% scaled-up \\   \cmidrule(lr){2-8}
         & ENS
         & 805.85 & 676.55 & 455.28 & 260.86 & 123.42 & 39.77  \\ \midrule
         
         \multirow{2}{*}{\begin{tabular}[c]{@{}l@{}}Load rationing\end{tabular}} 
         & Size
         & Baseline & 5\% Res. & 10\% Res. & 20\% Res. & 30\% Res. & 40\% Res. \\   \cmidrule(lr){2-8}
         & ENS
         & 805.85 & 641.79 & 528.97 & 308.03 & 154.10 & 95.72  \\ \midrule
         
         \multirow{3}{*}{\begin{tabular}[c]{@{}l@{}}Incentive-based\\ demand response\end{tabular}} 
         & Size
         & Baseline & 5\% Res. & 10\% Res. & 20\% Res. & 30\% Res. & 40\% Res. \\   \cmidrule(lr){2-8}
         & \multirow{2}{*}{\begin{tabular}[c]{@{}l@{}}ENS\end{tabular}}
         & \multirow{2}{*}{\begin{tabular}[c]{@{}c@{}}805.85\end{tabular}}
         & \multirow{2}{*}{\begin{tabular}[c]{@{}c@{}}670.30\\ $[$652.16, 681.30$]$\end{tabular}}
         & \multirow{2}{*}{\begin{tabular}[c]{@{}c@{}}563.65\\ $[$529.23, 594.05$]$\end{tabular}}
         & \multirow{2}{*}{\begin{tabular}[c]{@{}c@{}}431.56\\ $[$379.22, 473.33$]$\end{tabular}}
         & \multirow{2}{*}{\begin{tabular}[c]{@{}c@{}}357.63\\ $[$297.54, 414.47$]$\end{tabular}}
         & \multirow{2}{*}{\begin{tabular}[c]{@{}c@{}}322.48\\ $[$264.18, 372.48$]$ \end{tabular}}

        \\ &\\
        \bottomrule
    \end{tabular}
\end{adjustbox}
\caption{Quantified System Reliability with Single Demand Response Programs}
\label{tab:single}
\end{table}
Quantitative results show that the two deterministic mechanisms, interruptible load and residential load rationing, can help the ERCOT grid to avoid 75\% of forced load shedding by increasing the interruptible load to approximately 700\% or developing residential load rationing that curtails 27\% of residential demand. On the contrary, an incentive-based demand response program that covers 60\% of residences can only reduce forced load shedding by approximately 60\% on average. Echoing the ENS metric, the density distribution of forced load shedding with incentive-based demand response also reveals in detail its inefficiencies (Appendix A. 4).


\subsection{Interaction of Demand Flexibility with Mixing Mechanisms}
Considering the unusual share of the load sector in the February 2021 Texas extreme weather event and the decaying marginal effects of each single mechanism, it is imperative to facilitate multiple demand response programs that have different mechanisms and impact different load sectors to cooperatively exploit demand elasticity and improve system reliability under such extreme weather conditions. 
Using the same modeling for each mechanism, we examine multiple demand flexibility portfolios that consist of multiple mechanisms at different scales (see Section 3). In particular, we prioritize different mechanisms in the following order: incentive-based demand response, interruptible load, and residential load rationing.

\begin{figure}[h]
	\centering
	\includegraphics[width=0.4\textwidth]{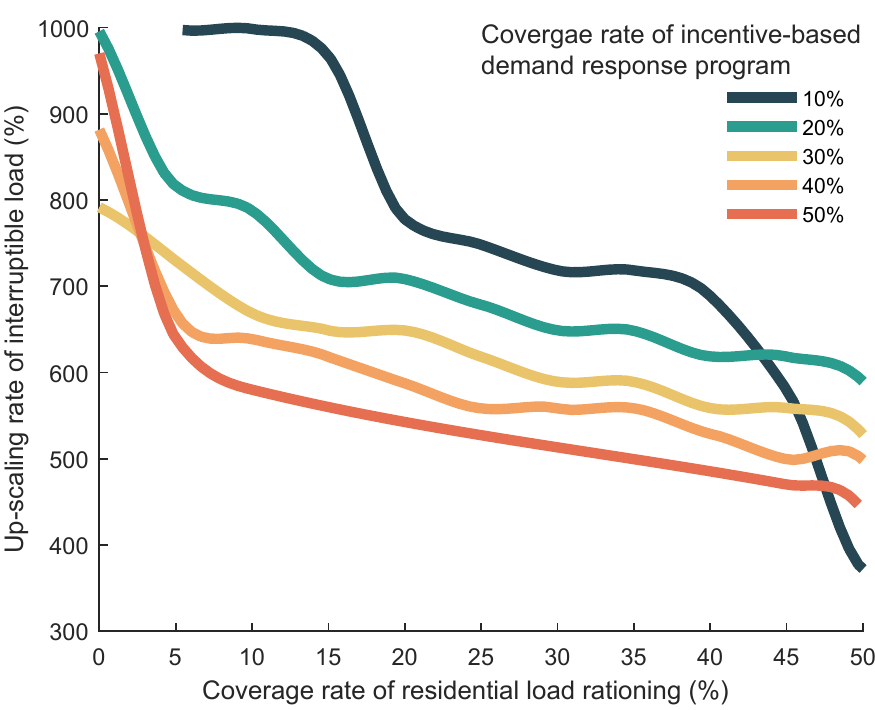}
	\caption{Required scales of interruptible loads and residential load rationing that can completely avoid the electricity outage given various scales of incentive-based demand response.}
	\label{fig:border_2D_monotone}
\end{figure}

We show that demand response portfolios can completely prevent power outages (Fig.~\ref{fig:border_2D_monotone}), as opposed to the observed performance of demand flexibility under the single demand response programs. Avoiding power outages through a demand response portfolio demonstrates the need and importance of deploying demand response response programs of different types in mixed load sectors.

\begin{figure}[h]
	\centering
	\includegraphics[width=1.0\textwidth]{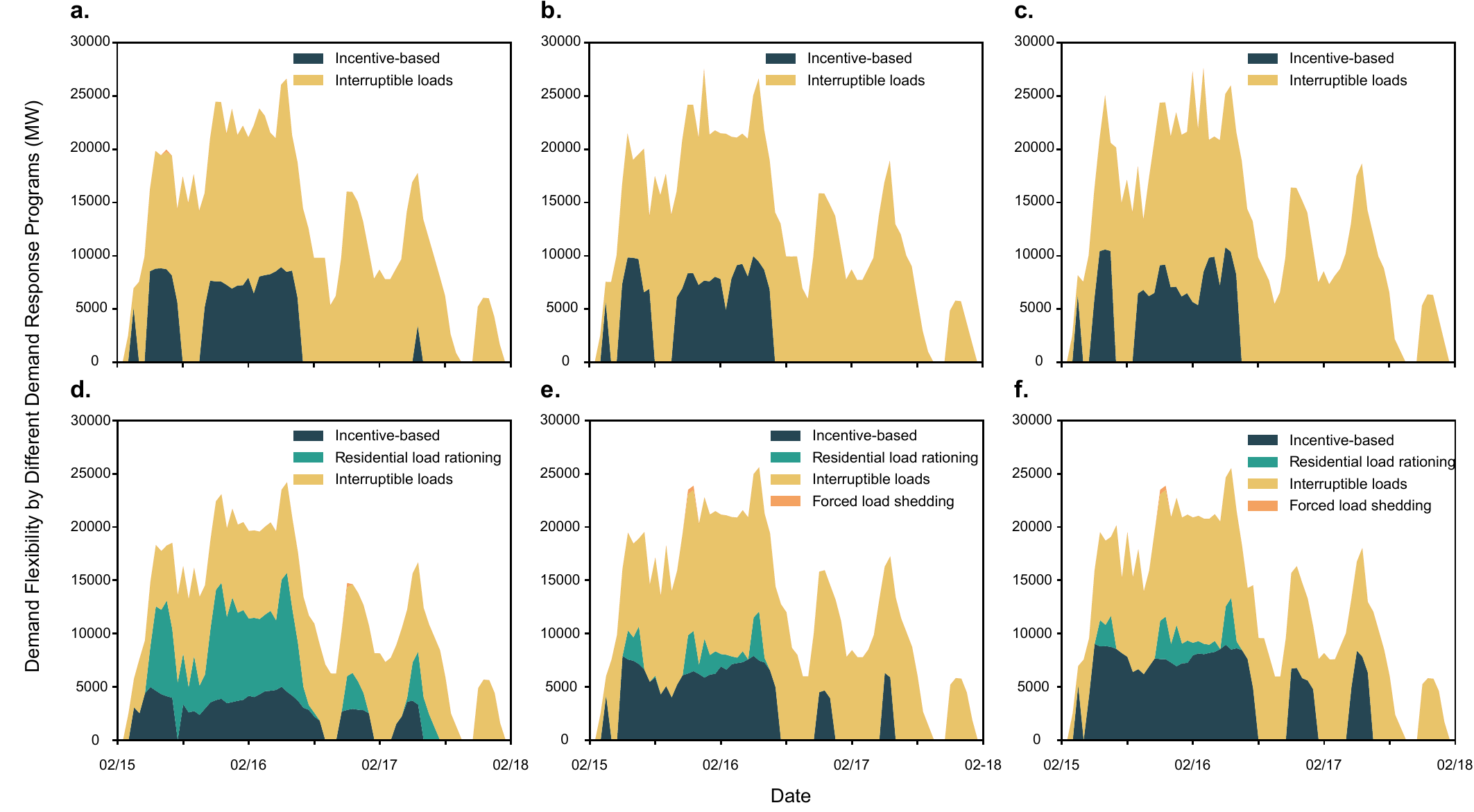}
	\caption{Demand flexibility provided by multiple mechanisms in different portfolios to completely avoid the electricity outage. The vertically stacked colored areas of time-varying thicknesses represent the real-time demand flexibility (in MW) provided by different mechanisms. a-c, Demand flexibility provided by incentive-based demand response (30$\%$, 40$\%$, and 50$\%$ coverage rate) and interruptible loads (790$\%$, 880$\%$, and 970$\%$ up-scaling rate), with no residential load rationing. d-f, Demand flexibility provided by incentive-based demand response (10$\%$, 20$\%$, and 30$\%$ coverage rate), interruptible loads (370$\%$, 590$\%$, and 530$\%$ up-scaling rate), and residential load rationing of 50$\%$ participation rate.}
	\label{fig:DR_portfolios_exmaples}
\end{figure}

Furthermore, the quantitative results reveal a complex interaction between different demand flexibility mechanisms (Fig.~\ref{fig:border_2D_monotone}). First, interruptible load and residential load rationing exhibit a clear complementary relationship, as reflected in the monotonic trend of each curve. Second, incentive-based demand response has complex impacts on the efficacy of both interruptible load and residential load rationing. Specifically, an incentive-based demand response program on a larger scale may require an even larger scale of the other two mechanisms to adequately avoid power outages, as reflected in curve intersections in Fig.~\ref{fig:border_2D_monotone}. To explain such counter-intuitive phenomena, we show in detail the time-varying demand flexibility provided by different mechanisms (Fig.~\ref{fig:DR_portfolios_exmaples} and Appendix A. 6). This phenomenon is conceptually similar to the effects of large amounts of renewable generation on the required generation reserve in an electric power grid. It is well known that a deeper penetration of renewable generation requires more firm generation capacity (such as thermal generation) to accommodate high uncertainty and variability. Similarly, incentive-based demand response with high variability on a larger scale may require additional demand flexibility of deterministic mechanisms to adequately prevent forced load shedding.



\section{Conclusions and Discussion}
In this study, we quantify the effects and interactions of heterogeneous demand flexibility mechanisms on the mitigation of power outages using a calibrated simulation model and data. We demonstrate that all single mechanisms have decaying marginal effects along with scale. The two deterministic programs, namely interruptible load and residential load rationing, can allow the Texas grid to avoid 75\% of forced load shedding by increasing the interruptible load to approximately 700\% or developing residential load rationing that covers 27\% of residences. In contrast, an incentive-based demand response program that covers 60\% of residential load can simply reduce forced load shedding by approximately 60\% on average. In addition, we identify portfolios of demand flexibility mechanisms that can completely avoid the February 2021 Texas power outage. Our results indicate that there are complex interactions between multiple mechanisms when included together in a demand response portfolio. We find that interruptible load and residential load rationing have a complementary relationship, while incentive-based demand response may have counterintuitive nonlinear impacts on the efficacy of the other mechanisms.

In summary, the quantitative results show how demand flexibility under multiple mechanisms could have helped to exploit multi-sector demand elasticity and improve the reliability of the Texas grid against the February 2021 Texas extreme weather, which also has reference value for other similar extreme events.  The fast-decaying marginal effects of demand flexibility require a balance between system reliability and economics. While developing a portfolio of multiple demand flexibility mechanisms that could potentially avoid power outages, the complex interaction between different mechanisms requires further investigation in critical scenarios in the future.
Regarding long-term planning, we emphasize the need to simultaneously facilitate multiple demand flexibility mechanisms that apply to different load sectors to robustly improve system reliability under unprecedented weather conditions, considering that unusual load sector shares may occur under extreme weather events and that decaying marginal effects are observed for individual mechanisms. For the practical effectiveness of a demand flexibility portfolio, it is critical to assess its performance under different possible scenarios of extreme events that have diverse patterns of multi-sector electricity demand.
Regarding the real-time operation of demand flexibility under extreme conditions, it is imperative for system operators to develop a more insightful and intelligent approach to allocating demand flexibility to different mechanisms and load sectors to maximize efficiency.

\section{Data and Code Availability}
Data, model and code that support the findings of this study are openly available at the following URL:~\url{https://github.com/tamu-engineering-research/DemandFlexibilityTexas}.

\Urlmuskip=0mu plus 1mu\relax  
\bibliography{ref}

\section*{Appendices}

\subsection*{A. 1: Brief Introduction of Demand Response Programs in Texas}
Demand flexibility is concretized by demand response programs of various forms that can change the energy consumption of end-use customers from normal patterns in response to certain commands or signals.~\cite{balijepalli2011review}
In most electricity markets in the United States, demand response providers (DRPs) are allowed to collect demand response commitments from customers and are treated as ``virtual generators” that can participate in real-time markets and/or provide operating reserves in the ancillary services markets.~\cite{Loadresource, Nyiso, Caiso}

Specifically, load resources in ERCOT can participate in the ancillary service market, which includes regulation up/down, non-spinning reserve service, and responsive reserve service (RRS). The bulk part of these services is attributed to the RRS, which can be automatically interrupted by under-frequency relays (UFRs) or manually cut-off within a ten-minute notice.~\cite{habib2021improving} During the most difficult hours of the Texas power outage, the maximum load reduction from load resources providing reserve was over 1,400 MW throughout the blackout event.~\cite{king2021timeline} 
Emergency Response Service (ERS) is another demand response program in the ERCOT, which will be activated during emergencies in the electric grid to avoid possible load shedding. There are two types of ERS with response times of 10 and 30 minutes procured by ERCOT four times a year.~\cite{drercot} 
Most demand response programs in ERCOT are focused on peak load reduction during summer. An essential example is the 4-Coincident Peak (4CP), where during each of the four months of June, July, August, and September, the single monthly system peak is set, and the 4CP charge for distribution
service providers are determined by calculating their load during ERCOT's monthly demand peak. 
Besides the demand response programs administered by ERCOT, there are also load management programs administered by Transmission and Distribution Service Providers (TDSP), which are an important source of demand flexibility. The three demand flexibility mechanisms considered in this paper cover most of these demand response programs. The ancillary services and the ERS are contractual interruptible loads, and 4CP is incentive-based demand response. One should note that although 4CP is a price-responsive demand response, in terms of structure and uncertainty, it is very similar to the coupon incentive-based demand response implemented in this paper. We obtain the available capacity of each demand response program in ERCOT from a presentation report,~\cite{drercot}  as follows. 
\begin{table}[h]
\centering
\begin{tabular}{|c|c|}
\hline
Demand response program & Capacity in ERCOT\\ \hline
Ancillary service market    & 2020 MW              \\ \hline
Emergency response service   & 1000 MW               \\ \hline
4-Coincident peak  & 2466 MW          \\ \hline
TDSP load management      & 300 MW          \\ \hline

\end{tabular}
\end{table}

\subsection*{A. 2: Details of Load Sector Share Estimation}
To improve the fidelity of our simulation and model different demand response proposals in richer detail, we further decompose the ERCOT weather-zone level total load data into 3 sectors: residential, commercial/industrial and municipal. We estimate a percentage share of the sectors in the total load of each weather zone, at each hour in February 2021 using the ERCOT backcasted actual load profile,~\cite{ERCOT_load_profiling} which contains normalized daily load profiles from each of the eight weather zones in ERCOT. The profile curves are produced using ERCOT load models which characterize the consumption pattern of loads in different sectors and with different sensitivity to weather and date-time variables. We first aggregate the consumption from all load model that belong to each sector and calculate the total daily variations for the February 2021 load profile data in each ERCOT weather zone. The sum of backcasted consumption of all models under each sector is the normalized to produce a timeseries of percentages. The hourly percentage timeseries are aligned with the historical hourly total consumption for each weather zone as shown in the table below, where $R_i$, $B_i$ and $O_i$ represent the percentage of consumption at hour $i$ for residential, business and other sector respectively. Note that, since each sector timeseries is normalized and scaled even in the raw ERCOT data, the percentages are of a hypothetical historic maximum MW capacity in each sector $[R_{max}, B_{max}, O_{max}]$ that is yet unknown (i.e. the MW consumption from the residential sector at hour $i$ should be $R_{max} * R_i$. For each hour, the estimated MW consumption from all 3 sectors, which is obtained by multiplying the total sector MW with the percentages ($R_i$, etc.), should sum up to the total historical MW consumption:
\begin{align}
R_i R_{max} + B_i B_{max} + O_i O{max} = P_i, \: \forall i
\end{align}
This formulation would allow us to use the data formatted in the table below to estimate the sector consumption through Non-Negative Least Square (NNLS), which gives the best estimation for the maximum sector MW consumption, $[R_{max}, B_{max}, O_{max}]$, from each sector and hence the hourly MW consumption from each sector.

\begin{table}[h]
\centering
\begin{tabular}{|c|c|c|c|c|}
\hline
Hour & Res. & Bus. & Oth. & Total MW \\ \hline
$T_1$        & $R_1$           & $B_1$        & $O_1$     & $P_1$             \\ \hline
$T_2$        & $R_2$           & $B_2$        & $O_t$     & $P_2$            \\ \hline
...        & ...             & ...          & ...       & ...             \\ \hline
$T_n$        & $R_n$           & $B_n$        & $O_n$     & $P_n$             \\ \hline
\end{tabular}
\end{table}

\subsection*{A. 3: Synthetic Texas Electric Grid Model}
\begin{figure*}[h]
	\centering
	\includegraphics[width=0.5\textwidth]{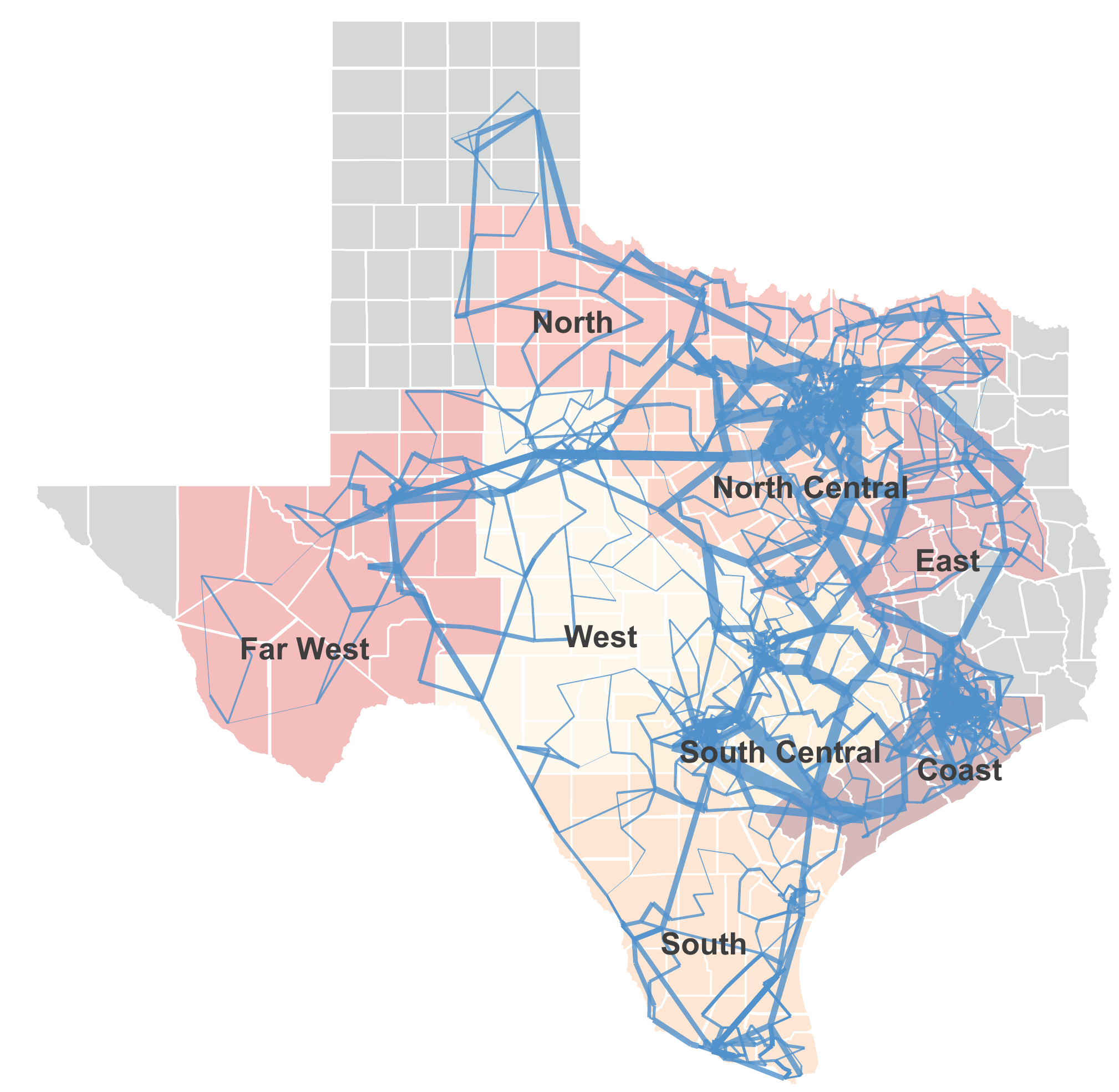}
	\caption{Visualization of synthetic simulation models over weather zones. It shows the topology of the synthetic Texas grid adapted from~\cite{wu2021open}, where the width is proportional to the transmission line capacity.}
	\label{fig:data_and_model}
\end{figure*}
\clearpage

\subsection*{A. 4: Forced load shedding density distribution}
\begin{figure*}[h]
	\centering
	\includegraphics[width=1\textwidth]{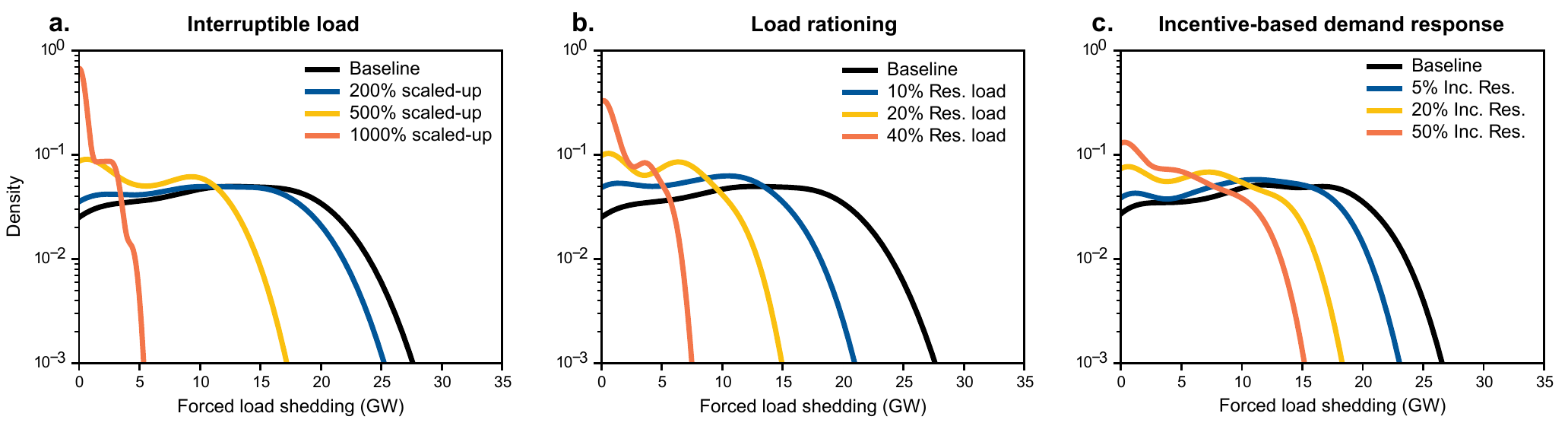}
	\caption{Density distributions of forced load shedding curves with demand flexibility of various mechanisms and scales}
	\label{fig:single-measure-density}
\end{figure*}
To present what-if blackout events clearly and comprehensively in an alternative way, we perform kernel density estimation (KDE) to calculate the density distribution of load shedding amount, which can serve as a complement to the ENS metric. ab, We observe that for interruptible load and load rationing, the peak of distribution becomes sharper and closer to zero along with the scale. c, Incentive-based demand response shows relatively lower sensitivity to the scale (Fig.~\ref{fig:single-measure-density}-c). Such observation also echos with the conclusion in the main text that merely developing incentive-based demand response cannot fully avoid electricity crisis due to the stochasticity and the limitation of inherent voluntary mechanism.

\subsection*{A. 5: Contour of Up-Scaling Multiples of Interruptible Load with Varying Scales of Incentive-Based Demand Response and Residential Load Rationing}
\begin{figure}[h]
	\centering
	\includegraphics[width=0.4\textwidth]{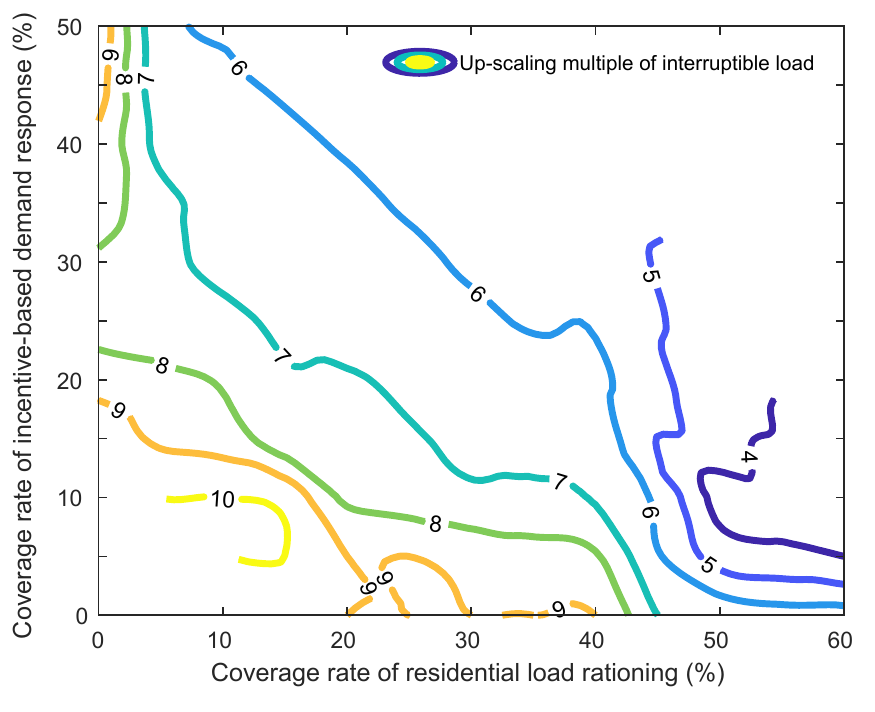}
	\caption{Contour of up-scaling multiples of interruptible load with varying scales of incentive-based demand response and residential load rationing, which can exactly avoid electricity outage.}
	\label{fig:contour}
\end{figure}
\clearpage

\subsection*{A. 6: Demand Flexibility Curves of Different Portfolios}
\begin{figure*}[h]
	\centering
	\includegraphics[width=0.9\textwidth]{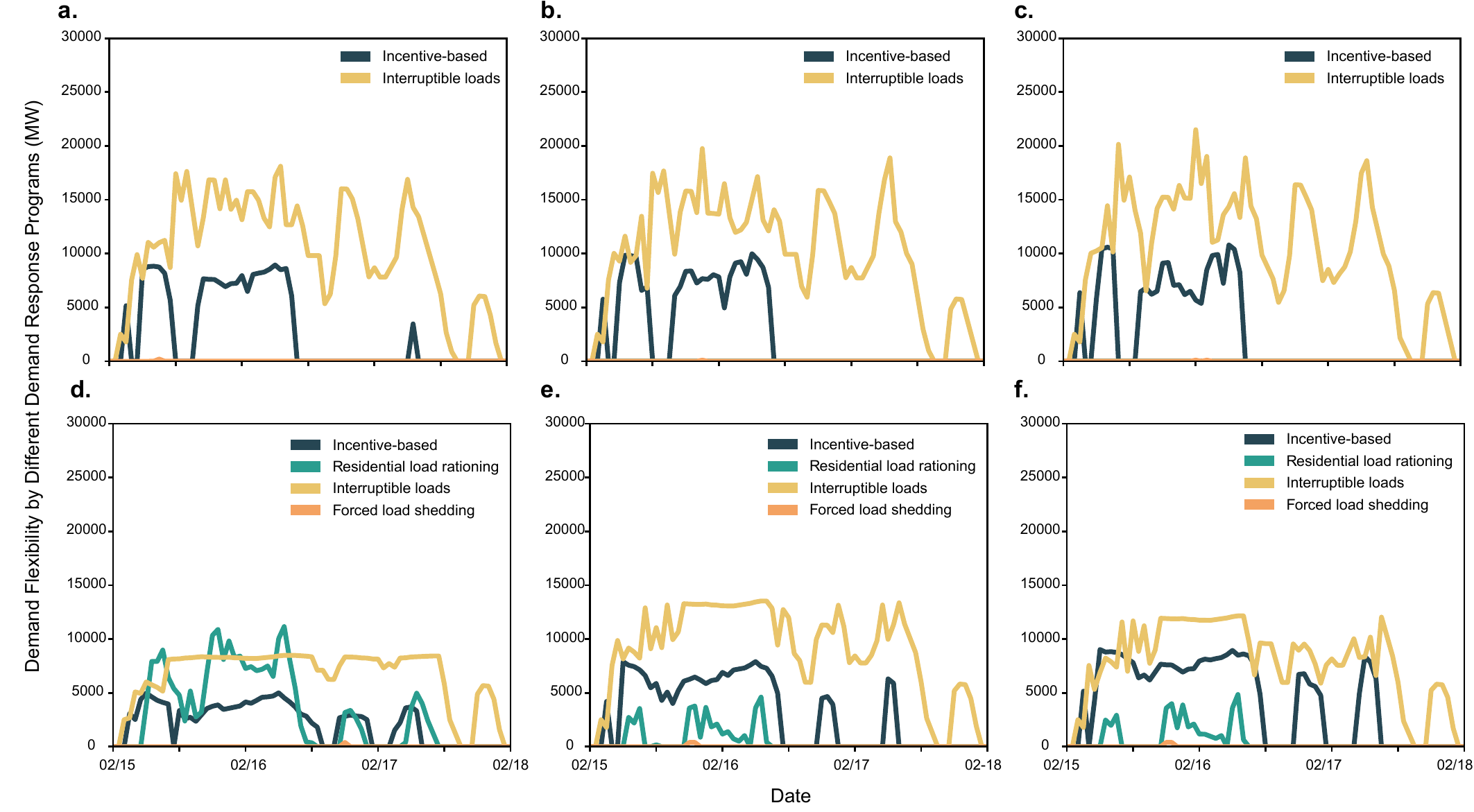}
	\caption{Demand flexibility provided by multiple mechanisms in different portfolios to exactly avoid electricity outage as supplementary information to Fig. 5 in the main text.}
	\label{fig:DR_portfolios_nonstacked_curves}
\end{figure*}

Sub-figure a-c, shows the demand flexibility provided by incentive-based demand response (30$\%$, 40$\%$ and 50$\%$ coverage rate) and interruptible loads (790$\%$, 880$\%$ and 970$\%$ up-scaling rate), with no residential load rationing. Sub-figure d-f shows the demand flexibility provided by incentive-based demand response (10$\%$, 20$\%$ and 30$\%$ coverage rate), interruptible loads (370$\%$, 590$\%$ and 530$\%$ up-scaling rate), and residential load rationing of 50$\%$ participation rate.

\clearpage

\end{document}